\documentclass{article}
\usepackage[a4paper,
            left=25mm,
            right=25mm,
            top=30mm,
            bottom=15mm,
            footskip=15mm,
            includefoot]{geometry}           
\usepackage{indentfirst}
\usepackage{fancyhdr}
\usepackage{main}
\usepackage{flushend}

\usepackage{times}
\usepackage{soul}
\usepackage{url}
\usepackage[hidelinks]{hyperref}
\usepackage[utf8]{inputenc}
\usepackage[small, format=plain, labelfont=it, textfont=it, font=normalsize]{caption}
\usepackage{graphicx}
\usepackage{mathastext}
\usepackage{amsmath}
\usepackage{amsfonts}
\usepackage{booktabs}
\usepackage{algorithm}
\usepackage{algorithmic}
\usepackage[sectionbib, numbers]{natbib}

\newcommand{\norm}[1]{\left\lVert#1\right\rVert}

\title{\vspace{-35pt}Synthetic Magnetic Resonance Images with Generative Adversarial Networks}

\author{
    \href{mailto:a.delplace@uq.net.au}{Antoine DELPLACE}
    \affiliations
    School of Information Technology and Electrical Engineering\\
    The University of Queensland, Qld., 4072, Australia
}

\fancypagestyle{plain}{%
  \fancyhf{}%
  \fancyfoot[C]{\thepage}%
}%

\begin{document}
\thispagestyle{plain}\enlargethispage{4\baselineskip}
\maketitle

\begin{abstract}
  Data augmentation is essential for medical research to increase the size of training datasets and achieve better results. In this work, we experiment three GAN architectures with different loss functions to generate new brain MRIs. The results show the importance of hyperparameter tuning and the use of mini-batch similarity layer in the Discriminator and gradient penalty in the loss function to achieve convergence with high quality and realism. Moreover, huge computation time is needed to generate indistinguishable images from the original dataset.
\end{abstract}

\section{Introduction}
Innovation in Computer Vision and Deep Learning is currently enhancing Medical Imaging research to improve diagnosis, image segmentation, and automated detection of specific cells or tissues. However, these recent methods require huge amount of data to train Neural Network models, and available resources in the Medical Imaging field are scarce.

On the other hand, new generative methods are today able to produce human faces with unprecedented quality and realism, giving access to unlimited datasets. The main idea relies on {\bf Generative Adversarial Networks (GAN)}, a dual architecture (see Subsection~\ref{gan}) which needs intensive hyperparameter tuning to reach high quality on the output images.

The following work presents a detailed comparison of three architectures to generate high quality Magnetic Resonance Images (MRIs), that can be used as input data for Neural Network training. Image quality, realism and diversity are studied with different hyperparameters along with computational efficiency. The three main methods presented here are: the original Deep Convolutional Generative Adversarial Network ({\bf DCGAN}), a Super Resolution Residual Network ({\bf SRResNet}) and a Progressive Generative Adversarial Network ({\bf ProGAN}).

\subsection{Related work}
The presented architectures are inspired by three different papers that move the quality of image generation forward.

\subsubsection{Generative Adversarial Networks}
\label{gan}
Goodfellow {\em et al.}~\cite{goodfellow-et-al:gan} introduces the original framework of GANs and presents the first promising results. The idea consists of training a first Neural Network, called {\bf Generator}, that tries to reproduce the input manifold as accurately as possible; and a second Neural Network, called {\bf Discriminator}, that tries to detect whether an input image comes from the input training data or the image collection created by the Generator.

Because the Generator and the Discriminator are competing against each other, the training is usually unstable and convergence is difficult to achieve. Hyperparameters need to be tuned to {\bf synchronize} the Generator and the Discriminator. Radford {\em et al.}~\cite{dcgan}, who introduced DCGANs, shows examples of generated images from LSUN (Large-scale Scene Understanding) and Imagenet-1k, a dataset of human faces.

\subsubsection{Super-Resolution GAN}
Since 2014~\cite{goodfellow-et-al:gan}, improvements in Deep Neural architectures have been made, enabling deeper and better performing networks. Ledig {\em et al.}~\cite{ledig-et-al:resgan} drew inspiration from Residual Network to create a Super-Resolution GAN that can upscale images with {\bf high frequency details}. The power of the architecture is its depth (with 16 residual blocks) leading to high accuracy but challenging configuration. Experiments on the Berkeley Segmentation Dataset (BSD100) show state-of-the-art performance although huge computing power is necessary.

\subsubsection{Progressive GAN}
Finally, Karras {\em et al.}~\cite{karras-et-al:progan} recently released a new technique to increase stability, speed and convergence during GAN training. The idea is to {\bf progressively increase the resolution} of the input and output images while adding deeper and deeper layers in the Generator and Discriminator. A {\bf minibatch standard deviation layer} is also added to the Discriminator to add diversity in the output space and prevent mode collapsing of the Generator. The paper shows incredible results in producing high quality faces using the {\sc Celeba-HQ} dataset, created for the experiments.

\subsection{Contribution}
The presented work defers from previous papers by analyzing different GAN architectures and several configurations dedicated to {\bf generating Magnetic Resonance Images} from a random latent space (noise). In particular, it goes beyond the work of Kazuhiro {\em et al.}~\cite{kazuhiro-et-al:mri-gan} by {\bf increasing the resolution} and {\bf benchmarking} convergence and quality of various methods.

The main contributions are:
\begin{itemize}
    \item successfully generating MRIs from noise with three different architectures: DCGAN, SRResNet, ProGAN
    \item the comparison of five loss functions: Original loss, LSGAN, WGAN, WGAN\_GP, DRAGAN
    \item the tuning of hyperparameters to improve convergence and quality
\end{itemize}

The different architectures and loss functions are presented in Section~\ref{method}. Then, a quantitative analysis on the Open Access Series of Imaging Studies (OASIS) dataset is provided in Section~\ref{experiments}. Finally, the paper concludes with a discussion in Section~\ref{discussion} and concluding remarks on possible future works in Section~\ref{conclusion}. The source code and results are available on the GitHub repository:\\ { \scriptsize \texttt{ \href{https://github.com/antoinedelplace/MRI-Generation}{https://github.com/antoinedelplace/MRI-Generation}}}.

\section{Method}
\label{method}
This Section covers the framework used to generate MRIs and the hyperparameter tuning process to increase convergence and stability.

\subsection{Tested architectures}
This work presents three main architectures for comparison: DCGAN, SRResNet and ProGAN.

\subsubsection{DCGAN}
The first architecture corresponds to a simple GAN using convolution layers. The detailed architecture is presented in Tables \ref{generator_architecture_dcgan} and \ref{discriminator_architecture_dcgan}. In particular, the input latent vector of the Generator is drawn from the distribution $\mathcal{U}(-1, 1)$, and the Leaky ReLU activation functions in the Discriminator have a slope of 0.2. Moreover, tensor weights are initialized with the distribution $\mathcal{N}(0, 0.02)$.

\begin{table}
    \centering
    \small
    \begin{tabular}{ccc@{ $\times$ }c@{ $\times$ }c}
        \toprule
        \textbf{Generator} & Act. & \multicolumn{3}{c}{Output shape} \\
        \midrule
        Latent vector & - & 256 & 1 & 1 \\
        Dense & BN+ReLU & 256 & 8 & 8 \\
        \midrule
        Conv Trans $5 \times 5$ & BN+ReLU & 256 & 16 & 16 \\
        Conv Trans $5 \times 5$ & BN+ReLU & 256 & 16 & 16 \\
        \midrule
        Conv Trans $5 \times 5$ & BN+ReLU & 256 & 32 & 32 \\
        Conv Trans $5 \times 5$ & BN+ReLU & 256 & 32 & 32 \\
        \midrule
        Conv Trans $5 \times 5$ & BN+ReLU & 256 & 64 & 64 \\
        Conv Trans $5 \times 5$ & BN+ReLU & 256 & 64 & 64 \\
        \midrule
        Conv Trans $5 \times 5$ & BN+ReLU & 128 & 128 & 128 \\
        Conv Trans $5 \times 5$ & BN+ReLU & 64 & 256 & 256 \\
        Conv Trans $5 \times 5$ & Tanh & 1 & 256 & 256 \\
        \bottomrule
    \end{tabular}
    \caption[table]{Generator Architecture of DCGAN}
    \label{generator_architecture_dcgan}
\end{table}
\begin{table}
    \centering
    \small
    \begin{tabular}{ccc@{ $\times$ }c@{ $\times$ }c}
        \toprule
        \textbf{Discriminator} & Act. & \multicolumn{3}{c}{Output shape} \\
        \midrule
        Input image & - & 1 & 256 & 256 \\
        \midrule
        Conv $5 \times 5$ & LReLU & 64 & 128 & 128 \\
        Conv $5 \times 5$ & BN+LReLU & 128 & 64 & 64 \\
        Conv $5 \times 5$ & BN+LReLU & 256 & 32 & 32 \\
        Conv $5 \times 5$ & BN+LReLU & 512 & 16 & 16 \\
        Conv $5 \times 5$ & BN+LReLU & 1024 & 8 & 8 \\
        \midrule
        Dense & LReLU & 1024 & 1 & 1 \\
        Dense & Sigmoid & 1 & 1 & 1 \\
        \bottomrule
    \end{tabular}
    \caption[table]{Discriminator Architecture of DCGAN}
    \label{discriminator_architecture_dcgan}
\end{table}

\subsubsection{SRResNet}
The second architecture is based on a Residual Neural Network with deeper layers. After a dense layer that rescales the input latent vector to the size $64 \times 16 \times 16$, the Generator is followed by 16 residual blocks (2 convolutions with $3 \times 3$ kernels). At the end, 4 upsampling blocks generate the $1 \times 256 \times 256$ output image by transferring filters into pixels, as described in Ledig {\em et al.}~\cite{ledig-et-al:resgan}. In the same way, the Discriminator is composed of 12 residual blocks separated by a downsampling convolution layer with $3 \times 3$ kernels and stride 2. The only dense layer is the one at the end of the Discriminator reducing the second to last layer of size $2048 \times 2 \times 2$ into one scalar. In addition, the Discriminator does not use any Batch Normalization in an attempt to avoid correlations within the batch. The 256-long input latent space is drawn from $\mathcal{N}(0, 1)$ before being normalized, and the tensor weights are initialized with the He Normal method.

\subsubsection{ProGAN}
The last architecture corresponds to a Progressive GAN as described in \cite{karras-et-al:progan}. Both networks are trained with increasing image resolutions from $4 \times 4$ to $256 \times 256$, with smooth image resolution transitions to progressively adapt the architecture. $5 \times 5$ kernels are used for convolutions and the latent input is a 512-long vector drawn from $\mathcal{N}(0, 1)$ before being normalized. Tensor weights are dynamically scaled at each iteration with the He Normal method. In addition, pixel normalization and mini-batch similarity layer are added to improve convergence and image diversity.

\subsection{Tested loss functions}
Because stability is difficult to reach when training GANs, different loss functions try to regularize and speed up the convergence. The following notations are use thereafter:
\begin{itemize}
    \item the input noise $\mathbf{z} \sim p_\mathbf{z}(\mathbf{z})$ (uniform or normalized normal)
    \item the generator output $G(\mathbf{z}) \sim p_g$
    \item the input data $\mathbf{x} \sim p_\mathrm{data}(\mathbf{x})$
    \item the ``probability'' $D(\mathbf{y})$, computed by the Discriminator, that $\mathbf{y}$ comes from $p_\mathrm{data}$ rather than $p_g$
\end{itemize}

\subsubsection{Original Loss}
Equation \ref{loss_gan} is the original loss introduced in \cite{goodfellow-et-al:gan} inspired by binary cross entropy loss but slightly modified to prevent vanishing gradient during backpropagation.
\vspace{-6pt}
\begin{align}%
\label{loss_gan}%
    L_D^\mathrm{GAN} &= -~\mathbb{E} \bigg[ \log \big( D(\mathbf{x}) \big) \bigg] - \mathbb{E} \bigg[ \log \big( 1-D(G(\mathbf{z})) \big) \bigg] \nonumber\\
    L_G^\mathrm{GAN} &= -~\mathbb{E} \bigg[ \log \big( D(G(\mathbf{z})) \big) \bigg]
\end{align}%

\subsubsection{LSGAN}
LSGAN (Equation \ref{loss_lsgan}) is another loss function that tries to reduce mode collapsing and vanishing gradient.
\vspace{-12pt}
\begin{equation}%
\label{loss_lsgan}%
\begin{split}%
    L_D^\mathrm{LSGAN} &= \mathbb{E} \bigg[ \big( D(\mathbf{x}) - 1 \big)^2 \bigg] + \mathbb{E} \bigg[ D(G(\mathbf{z}))^2 \bigg]\\
    L_G^\mathrm{LSGAN} &= \mathbb{E} \bigg[ \big( D(G(\mathbf{z})) - 1 \big)^2 \bigg]
\end{split}%
\end{equation}%

\subsubsection{WGAN}
In order to increase stability and convergence, WGAN (Equation \ref{loss_wgan}) replaces the original Jensen-Shannon divergence by the Wasserstein distance, a continuous function where the gradient is more easily computed.
\begin{equation}%
\label{loss_wgan}%
\begin{split}%
    L_D^\mathrm{WGAN} &= -~\mathbb{E} \big[ D(\mathbf{x}) \big] + \mathbb{E} \big[ D(G(\mathbf{z})) \big]\\
    L_G^\mathrm{WGAN} &= -~\mathbb{E} \big[ D(G(\mathbf{z})) \big]
\end{split}%
\end{equation}%

\subsubsection{WGAN\_GP}
In addition to the WGAN loss, a gradient penalty can be introduced to avoid exploding gradient. $\lambda$ is a hyperparameter to balance the penalty and $\alpha \sim \mathcal{U}(0, 1)$ is a random parameter that combines the real and fake image.
\begin{equation}%
\begin{split}%
    L_D^\mathrm{WGAN\_GP} &= L_D^\mathrm{WGAN} + \lambda \mathbb{E} \bigg[ \big( \norm{\nabla D(\mathbf{x_m})} - 1 \big)^2 \bigg]\\
    \mathbf{x_m} &= \alpha \cdot \mathbf{x} + (1-\alpha) \cdot  G(\mathbf{z})
\end{split}%
\end{equation}%

\subsubsection{DRAGAN}
With the same idea, DRAGAN (Equation \ref{loss_dragan}) introduces a gradient regularization to avoid local minima and mode collapsing. $\lambda$ is a hyperparameter and $\alpha \sim \mathcal{U}(0, 1)$ combines the real image with $\mathbf{x_p} \sim \mathcal{U}(0, 0.5\sigma_\mathbf{x})$, a pixel-scaled random noise.
\begin{equation}%
\label{loss_dragan}%
\begin{split}%
    L_D^\mathrm{DRAGAN} &= L_D^\mathrm{GAN} + \lambda \mathbb{E} \bigg[ \big( \norm{\nabla D(\mathbf{x_m'})} - 1 \big)^2 \bigg]\\
    \mathbf{x_m'} &= \alpha \cdot \mathbf{x} + (1-\alpha) \cdot  \mathbf{x_p}
\end{split}%
\end{equation}%

\subsection{Tricks to improve GAN performance}
The presented architectures are the result of a lot of trials and errors and a deep literature investigation. During the first steps of the experiments, it had been noticed that using a {\bf one-sided smoothing label} (ie training both networks with $D(\mathbf{y}) = 0.9$ instead of 1.0) gave better visual results. Moreover, most of the tested networks have a Discriminator too strong. That is why the {\bf Generator/Discriminator rate} $r$ is often greater than 1 ($r=3$ for DCGAN, $r=2$ for SRResNet and $r=1$ for ProGAN).

\section{Experiments}
\label{experiments}
The different models are trained on a Tesla K80 GPU with a batch size of 64 images. 60 epochs are used except for the ProGAN which needs 20 epochs for each change in resolution and after each block transition ($20*13=260$ epochs). The Adam optimizer performs the minimization of the loss function with $l_r=0.0002$ and $\beta_1=0.5$ for DCGAN and SRResNet, but with $l_r=0.001$ and $\beta_1=0$ for ProGAN.

\subsection{Training dataset}
The training dataset comes from the Open Access Series of Imaging Studies (OASIS). It is composed of 11328 brain MRIs of size $256 \times 256$, scaled to $[-1, 1]$.

\subsection{Evaluation measures}
Two main quality measures need to be estimated: the image {\bf realism} and the generated manifold {\bf diversity}. To do so, a Principle Component Analysis (PCA) is performed over the data distribution to obtain 16 orthogonal vectors that represent the main variations of the input manifold (55 \%).

Realism $\rho$ is calculated by projecting $N=11328$ generated images $\mathbf{G}$ on the 16 covariance matrix eigenvectors $\mathbf{E}_i$ and retrieving the mean of the cosine similarity vector norm.
\begin{equation}%
\rho = \frac{1}{N}\sum_{\mathbf{G}} \sqrt{\sum_{i=1}^{16} (\mathbf{G}\cdot\mathbf{E}_i)^2 }
\end{equation}%

Diversity is evaluated through 2 measures: the total variation $\sigma$ of the generated set and the number $\delta$ of covariance matrix eigenvalues which are greater than 1\% of $\sigma$.

$\rho$ is meant to detect unrealistic images, $\sigma$ detects if the images have too few variations whereas $\delta$ tracks mode collapsing.

Finally, an estimation of how the model overfits by remembering all training images is performed by visualizing images generated from the interpolation between two random latent vectors.

\subsection{Results}
\begin{figure*}
\centering
	\frame{\includegraphics[decodearray={1 0}, width=0.25\textwidth]{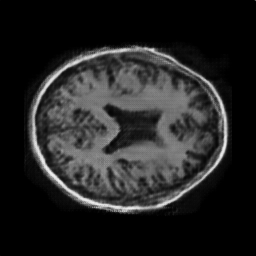}}%
	\frame{\includegraphics[decodearray={1 0}, width=0.25\textwidth]{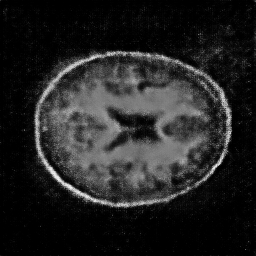}}%
	\frame{\includegraphics[decodearray={1 0}, width=0.25\textwidth]{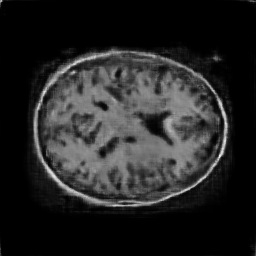}}%
	\frame{\includegraphics[decodearray={1 0}, width=0.25\textwidth]{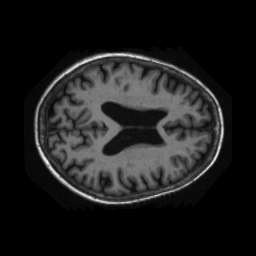}}%
    \caption[figure]{Generated images, from left to right: DCGAN, SRResNet, ProGAN, Original}
    \label{plots}
\end{figure*}

The Figure \ref{plots} shows a generated image for each main architecture. It can be noticed that they are all close to the original distribution, even if they seem a bit blurry compared to the ground truth. The quality of the images can be improved by increasing the number of epochs, but requires more computation time.

The Table \ref{evaluation} summarizes the evaluation measures for each architecture. DRAGAN and WGAN\_GP are the only loss functions that allow stabilized training. DCGAN performs better with DRAGAN whereas SRResNet and ProGAN converge only with respectively DRAGAN and WGAN\_GP loss. Finally, DCGAN performs better than SRResNet or ProGAN, and SRResNet is the fastest to train (30 hours) compared to DCGAN (45 hours) and ProGAN (58 hours).
\begin{table}
    \centering
    \small
    \begin{tabular}{ccccc}
        \toprule
        \textbf{Model} & \textbf{Loss} & $\rho$ & $\sigma$ & $\delta$ \\
        \midrule
        Training dataset & - & 0.739 & 409 & 15 \\
        \midrule
        DCGAN & DRAGAN & 0.718 & 352 & 15 \\
        SRResNet & DRAGAN & 0.678 & 329 & 16 \\
        ProGAN & WGAN\_GP & 0.601 & 331 & 13 \\
        \bottomrule
    \end{tabular}
    \caption[table]{Realism and Diversity evaluation}
    \label{evaluation}
\end{table}

\section{Discussion}
\label{discussion}
The performed experiments reinforce evidence that GANs are {\bf difficult to train} and are {\bf sensitive to small changes} in the hyperparameters or architecture. However, it has been shown that using a mini-batch similarity layer in the Discriminator (used in DCGAN and ProGAN) and controlling the gradient norm (used in DRAGAN and WGAN\_GP) are essential to stabilize the training. On the contrary, the use of noise in the discriminator or the choice between uniform or normalized Gaussian latent input do not seem to have any impact on the quality or realism of the results.

\section{Conclusion and future work}
\label{conclusion}
To conclude, GANs can be used to increase MRI datasets (data augmentation) and thus enable more advance training for neural networks. However, {\bf huge computation time} is needed to achieve high quality and realism, and produce generated images indistinguishable from the original dataset.

Future work must be focused on training new architecture (like StyleGAN) or performing {\bf 3D GANs} that can generated 3D images without running out of memory.

\section*{Acknowledgments}
I would like to thank Dr.~Shakes Chandra for his useful insight on this project and for the helpful supervision he gave me when completing my Master thesis.

\bibliographystyle{mybibstyle}
\bibliography{main}

\section*{Biography}
Antoine Delplace is a Master student pursuing a double degree at the University of Queensland and at Ecole Centrale Paris in Software Engineering. His research focuses on Machine Learning and Deep Neural Networks. He will begin a PhD degree at the University of Queensland in January 2020.

\end{document}